\documentclass[12pt,a4paper]{article}
\usepackage{amssymb,amsmath,amscd,epsfig}
\title{Oscillations of  neutrinos produced by a beam of electrons}
\author{
A.D. Dolgov\thanks{e-mail: dolgov@itep.ru}\hspace*{2mm}$^{\rm
a,b,c}$, L.B. Okun\thanks{e-mail: okun@itep.ru}\hspace*{2mm}$^{\rm
a}$, M.V. Rotaev\thanks{e-mail:
mrotaev@mail.ru}\hspace*{2mm}$^{\rm a,d}$, and
M.G. Schepkin\thanks{e-mail: schepkin@itep.ru}\hspace*{2mm}$^{\rm
a}$\\[5mm]
${\rm ^a}$ {\small\it Institute of Theoretical and
Experimental Physics}\\ {\small\it 117218, B.Cheremushkinskaya 25,
Moscow, Russia}\\
${\rm ^b}$ {\small\it INFN, Ferrara 40100,
Italy} \\
${\rm ^c}$ {\small\it ICTP, Trieste, 34014,
Italy} \\
${\rm ^d}$ {\small\it Moscow Physics and Technology
Institute}}

\date{}

\begin{document}

\newcommand{\x}{\mathbf{x}}
\newcommand{\y}{\mathbf{y}}
\newcommand{\p}{\mathbf{p}}
\newcommand{\K}{\mathbf{k}}
\newcommand{\n}{\mathbf{n}}
\newcommand{\q}{\mathbf{q}}
\newcommand{\e}{\mathbf{e}}
\newcommand{\V}{\mathbf{v}}

\maketitle

\begin{abstract}

We analyze a thought  neutrino oscillation experiment in which a beam of
neutrinos is produced by electrons colliding with atomic nuclei of a target. 
The neutrinos are detected by observing charged leptons, which are produced 
by neutrinos colliding with nuclei of the detector. We consider the case 
when both the target and detector nuclei have finite masses. (The case of 
infinitely heavy nuclei was considered in the literature earlier.)

\end{abstract}

\section{Introduction}

Despite  an impressive number of theoretical papers published
during the last 40-50 years, the phenomenology and description of
neutrino oscillations is still a subject of heated debates. In
particular, there is no consensus on the assumptions of equal
energies or equal momenta of the three neutrino mass eigenstates
$\nu_j$, $j=1,2,3$. 

The equal momenta scenario was introduced in a pioneering paper 
on neutrino oscillations by Gribov and Pontecorvo \cite{GribovPont},
used by  Fritzsch and Minkowski \cite{Fritzsch} and then
by many other authors.
The equal energies scenario was presented by Kobzarev et al. \cite{KMOS}, 
who considered  all three virtual neutrinos produced by a monochromatic
beam of electrons colliding with  infinitely heavy nuclei 
($M \to \infty$). Since the recoil energy of such nuclei is zero, all three 
neutrinos have equal energies, the same as the energy of the electron. 
Stodolsky \cite{Stod}, Lipkin \cite{Lipkin} and Vysotsky \cite{Vys} presented 
general arguments in favor of equal energy scenario for realistic thought 
experiments. Still, in the most recent and authoritative review of particle 
physics in the contribution by Kaiser \cite{PDG4} neutrinos oscillations are 
discussed on the basis of equal momentum scenario. Note that the same attitude 
one can find in his previous review \cite{PDG2}, while in 2000 \cite{PDG0} 
both equal energy and equal momentum scenarios were considered on the same 
footing (all this - in the oversimplified  "neutrino 
 plane wave approximation"). 
 
In refs. \cite{GribovPont},\cite{Fritzsch},\cite{PDG4},\cite{PDG2},
\cite{PDG0} plane wave free neutrinos traveling from the 
production point $A$ to the detection point $B$  were considered
without discussing their progenitors. We will refer to such
descriptions as reduced ones. Following the argument of 
ref. \cite{Vys}  it is evident that in the "reduced approach" 
only the equal energy scenario is self-consistent. Otherwise
the neutrinos are produced at the point $A$ not in a given
flavor state, but in a state whose flavor oscillates with time.

In ref. \cite{KMOS} the progenitor (electron) was described by 
a plane wave. There  exists a vast literature in which both
the progenitor and offspring particles are described not by
plane waves but by wave packets (see review by Beuthe 
\cite{MB}). In the present paper our attention is concentrated
on the incoming and outgoing particles. We will find that 
depending on the properties of these external particles
 both  differences of energies and
momenta of different neutrino mass eigenstates are 
non-vanishing, but the energy difference is much smaller 
than the momentum difference when the energy transfer to
the target nucleus is small.

In this note we are going to consider a more realistic situation
than in ref. \cite{KMOS}, namely, when the beam of electrons is not
monochromatic (it is described by a finite-size wave packet), and the mass 
of the target nucleus is finite. Now the recoil energy  of the nucleus cannot 
be neglected. The detection of neutrino occurs when it interacts with another 
nucleus (in detector).
When a nucleus is in a crystal it is described by a wave function with 
a characteristic momentum spread about 1 keV and vanishing mean momentum. 
When the nucleus is in gas, its momentum is not vanishing, while momentum 
spread is smaller. (Note that even for an infinitely heavy nucleus in crystal 
the spread of the momentum is non-zero). 

We will prove that in the case of finite masses of
 nuclei $A$ or $B$ neutrino oscillations disappear in the limit of the 
vanishing momentum spread of the electron wave packet (plane wave limit).

The structure of this paper is the following. We introduce "little donkey" 
diagram with a virtual neutrino propagating between production and detection 
points in section 2. The amplitude is derived and analyzed 
in section 3. The expression of the phase difference responsible 
for oscillations is discussed in section 4.
The probability of neutrino oscillations and their suppression is
discussed in section 5. Section 6 is devoted 
to  concluding remarks.

\section{Probability and "little donkey" diagram}

Let us consider interaction of an electron $e$ with a nucleus $A$
of mass $M_A$ in a target. 
Neutrino produced in this interaction collides
later on in a detector with a nucleus $B$ of mass $M_B$ and produces
a charged lepton $l$. As a result the whole process looks like
$e+A+B \rightarrow l+C+D$, (see Fig. 1).

The electronic neutrino $\nu_e$ produced  on nucleus $A$ is a
superposition of three neutrino mass eigenstates:
$\nu_e = \sum_i U_{e i}\nu_i$ where $\nu_i$ is the state with mass
$m_i$. Each mass eigenstate propagates independently between nuclei
$A$ and $B$. Interaction with $B$ results in projection of the three
neutrino propagating states on the state $\nu_l= \sum_i U_{l i}\nu_i$.
Here
 $U$ is the unitary mixing matrix, the first and second indices
of which  denote respectively flavor and mass eigenstates.  

\begin{figure}
\centerline{\epsfig{file=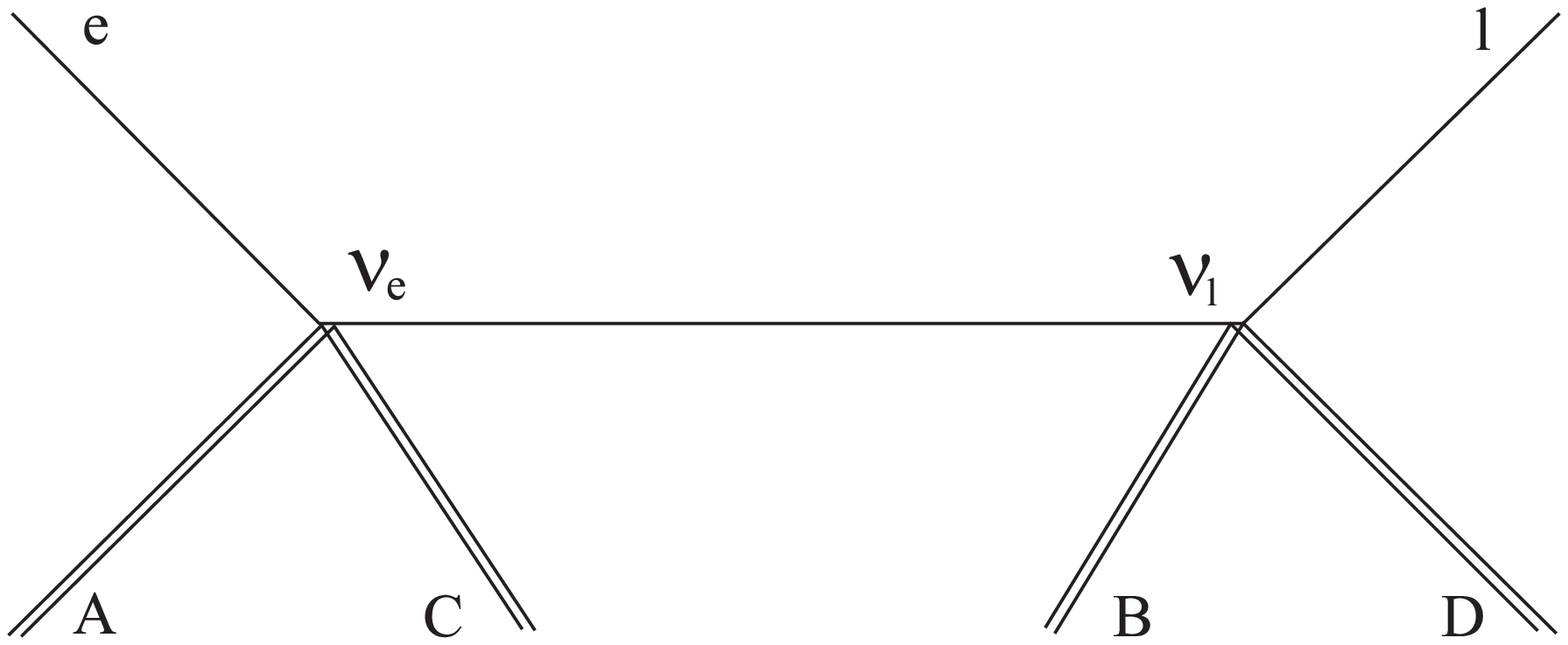, width=10.5cm, height=5cm}}
\caption{Little donkey diagram}
\label{}
\end{figure}

 In oscillation experiments the nuclei $C$ and $D$ are not
 registered, while the energy and momentum of the lepton $l$
 are measured with low precision. Thus the probability of
 the whole process is obtained by integration 
 of the amplitude squared, possibly weighted with the detector
resolution function, 
over $d\p_l d\p_C d\p_D$

\begin{multline}
\label{P_{exp}}
P(\p_e,\p_{A,B},M_{A,B},\x_{A,B},\sigma_{A,B})
=\int \sum_{i,j}U_{e i} U_{l i}^* U_{e j}^* U_{l j} \times \\ \times
P_{ij}(\p_{e,l},\p_{A,B},M_{A,B},\x_{A,B},\sigma_{A,B},\p_{C,D},E_{C,D})
\frac{d\p_l}{2E_l}\frac{d\p_C}{2E_C}\frac{d\p_D}{2E_D}.
\end{multline}
Here $P_{i j}=\mathcal{A}_i^{\star}\mathcal{A}_j$,
where $\mathcal{A}_i$ is the amplitude for a given neutrino state
with mass  $m_i$; symbols like $\p_{A,B}$ denote $\p_A$, $\p_B$. Let us 
express $\mathcal{A}_j$ in terms of the wave functions of all interacting 
particles: $e, A, B, C, D, l$ and neutrino propagator $G(x_1, x_2)$ :
\begin{multline}
\label{Ampl}
\mathcal{A}_j(\p_{e,l},
\p_{A,B},M_{A,B},\x_{A,B},\sigma_{A,B},\p_{C,D},E_{C,D}) =
 \int d\x_1 d\x_2 dt_1 dt_2 \times \\ \times
\psi_e(\x_1,t_1)\psi_{A}(\x_1,t_1)\psi_{B}(\x_2,t_2) G_j(\x_1,t,\x_2,t_2)
\psi_{l}^{\ast}(\x_2,t_2)\psi_{C}^{\ast}(\x_1,t_1)\psi_{D}^{\ast}(\x_2,t_2).
\end{multline}

In ref. {\cite{KMOS}} electron $e$ and lepton $l$ were 
described by plane waves:

\begin{equation}
\label{leptons}
\begin{array}{rcl}
\psi_e(\x_1,t_1) & = & e^{i\p_e \x_1-iE_e t_1}, \\
\psi_{l}(\x_2,t_2) & = & e^{i\p_l \x_2-iE_l t_2}, \\
\end{array}
\end{equation}
while nuclei $A$ and $B$ were described by $\delta$-functions in 
configuration space
{\footnote {The equality signs in equations throughout the paper
should be taken "with a grain of salt" because we omit some
obvious normalization factors. This makes the formulas easier
to read without influencing the physical results, e.g.
the ratio of oscillating 
terms to the non-oscillating ones.}}.

Now we assume that nuclei $A$ and $B$ are described by the
finite-size wave packets: 
\begin{equation}
\label{InNucl's}
\begin{array}{rcl}
\psi_{A}(\x_1,t_1) & = & \int e^{-\frac{(\q_A-\p_A)^2}{2\sigma_A^2}}
e^{i\q_A(\x_1-\x_A)-iE_A(\q_A)(t_1 - t_A)} d\q_A~, \\
\psi_{B}(\x_2,t_2) & = & \int e^{-\frac{(\q_B-\p_B)^2}{2\sigma_B^2}}
e^{i\q_B(\x_2-\x_B)-iE_B(\q_B) (t_2 - t_B)} d\q_B~,
\end{array}
\end{equation}
where $\p_A$ and $\p_B$ are the central momenta of the wave packets of 
nuclei $A$ and $B$ respectively, while $\q_A$ and $\q_B$ are the 
corresponding running momenta.  In eqs. (\ref{InNucl's}) and below
we use Gaussian wave packets. The main features of our results do
not depend upon the specific form of the packets.
In our subsequent publication we are
going to take for them a general form.

The wave functions of nuclei satisfy Klein-Gordon equation 
as we consistently neglect the spins of all particles.
All external particles are assumed to be free and hence 
their momenta satisfy the on-mass-shell condition:

\begin{equation}
\label{Klein-Gordon}
E_n(\q_n) = \sqrt{\q_n^2+M_n^2}~,
\end{equation}
where the index $n$ denotes $A$, $B$ and $e$.
The same is true also for outgoing particles. 

Since in  
neutrino oscillation experiments the final nuclei $C$ and $D$ 
are not registered, one can choose  for their wave functions
any basis in the Hilbert space, and then 
integrate the probability over all Hilbert space.
In what follows we take plane waves as the complete 
 and orthogonal 
set of wave functions:

\begin{equation}
\label{OutNucl's}
\begin{array}{rcl}
\psi_{C}(\x_1,t_1) & = & e^{i\p_C \x_1-iE_C (t_1-t_A)}~, \\
\psi_{D}(\x_2,t_2) & = & e^{i\p_D \x_2-iE_D (t_2-t_B)}~,
\end{array}
\end{equation}
where $E_C^2=\p_C^2+M_C^2$ and $E_D^2=\p_D^2+M_D^2$.
We think that description of outgoing particles by wave packets 
in the amplitude is not a
consistent procedure, because generally wave packets do not have
orthogonality property, neither they form a complete set of functions.

Here we would like to touch upon a subtle point. In a more or less realistic 
thought experiment the target and the detector are solids, therefore the 
on-mass-shell condition 
for nuclei is only an approximation. This approximation 
seems to be reasonable for ordinary matter, where nuclei are weakly bound.

Instead of the plane wave of ref. \cite{KMOS}, the wave function of the 
electron is described  now  
by the one dimensional wave packet with definite 
direction $\e = {\p_e}/{|\p_e|}$  of the beam:
\begin{equation}
\psi_{e}(\x_1,t_1) = \int e^{-\frac{(\q_e-\p_e)^2}{2\sigma_e^2}}
e^{i\q_e \cdot (\x_1-\x_e)-iE_e(\q_e)(t_1 - t_e)} d\q_e.
\end{equation}
Here and in the following:
\begin{equation}
\label{dq_e}
d\q_e \equiv \delta(\e - \frac{\q_e}{|\q_e|}) q_e^2 dq_e d\Omega_e,
\end{equation}
where $\Omega_e$ is the corresponding solid angle, and $q_e \equiv |\q_e|$.
We will show  below that the oscillation terms vanish when 
$\sigma_e$ tends to zero. We choose the one-dimensional
packet  only because of technical simplicity.
The result can be obtained in a more general case.

\section{Neutrino Green function and the amplitude}

Following  ref. \cite{KMOS} we replace the neutrino Green 
function with the propagator of a scalar particle of  
mass $m_j$, where $j$ numerates 
neutrino mass eigenstates, $j=1,2,3$; it is clear that fermionic nature of 
the neutrino (as well as of $e$ and $l$) is not essential in the problem. Thus

\begin{equation}
\label{prop}
G_j(\x,t)=\frac{1}{(2\pi)^4} \int
\frac{e^{-i\omega t+i\K\x}}
{\omega^2-\K^2-m^2_j+i\varepsilon}d\K d\omega~.
\end{equation}

For each $\nu_j$ the amplitude of the process is written as

\begin{multline}
\label{Ampl'}
\mathcal{A}_j = \int 
e^{-\frac{(\p_e-\q_e)^2}{2\sigma_e^2}}
e^{i\q_e (\x_1-\x_e) -iE_e (t_1-t_e)}d\q_e
e^{-\frac{(\q_A-\p_A)^2}{2\sigma_A^2}} e^{i\q_A(\x_1-\x_A)-iE_A(\q_A) 
(t_1-t_A)} d\q_A
\times \\ \times
G_j(\x_1-\x_2,t_1-t_2) \times
e^{-\frac{(\q_B-\p_B)^2}{2\sigma_B^2}} e^{i\q_B(\x_2-\x_B)-iE_B(\q_B) 
(t_2-t_B)} d\q_B
e^{-i\p_l \x_2+iE_l (t_2-t_B)} 
\\ \times
e^{-i\p_C \x_1+iE_C (t_1-t_A)}e^{-i\p_D \x_2+iE_D (t_2-t_B)} 
d\x_1 d\x_2 dt_1 dt_2 
= \\ = 
\int e^{-\frac{(\p_e-\q_e)^2}{2\sigma_e^2}}
e^{i\q_e (\x_1-\x_e)+iE_e (t_e-t_A)} 
e^{-\frac{(\q_A-\p_A)^2}{2\sigma_A^2}}
e^{i\q_A(\x_1-\x_A)} e^{-\frac{(\q_B-\p_B)^2}{2\sigma_B^2}} 
\times \\ \times 
e^{i\q_B(\x_2-\x_B)}
\frac{1}{4\pi|\x_1-\x_2|}e^{-ik_j|\x_1-\x_2| + i\omega \cdot (t_A-t_B)}
e^{-i\p_l \x_2} e^{-i\p_C \x_1-i\p_D \x_2} d\x_1 d\x_2 
\times \\ \times
\delta(E_e + E_A(\q_A)+E_B(\q_B) - E_C - E_l- E_D) d\q_e d\q_A d\q_B~,
\end{multline}
where we use $dt_1 dt_2=\frac{1}{2}d(t_1+t_2)d(t_1-t_2)$, and 

\begin{equation}
\int e^{-i\omega(t_1-t_2)}G_j(\x_1-\x_2,t_1-t_2)d(t_1-t_2) =
\frac{1}{4\pi|\x_1-\x_2|}e^{i\sqrt{\omega^2-m_j^2}|\x_1-\x_2|}~.
\end{equation}
The parameter $\omega$ is defined by

\begin{equation} 
\label{omega-p-dep.}
\begin{array}{rcccccl}
\omega & \equiv & \omega(\q_A,\q_B) & = & E_e+E_A(\q_A)-E_C & = & 
E_l+E_D-E_B(\q_B),
\end{array}
\end{equation}
and

\begin{equation}
\label{kappa}
k_j \equiv \sqrt{\omega^2-m_j^2}.
\end{equation}
Though $k_j$ looks like a three-momentum, in fact, it is a short-hand 
notation, usually arising in description of propagation of spherical waves with definite energy.

The integration over $d(t_1+t_2)$ in eq.(\ref{Ampl'})  
gives $\delta$-function leading to  energy conservation.

The further analysis of the problem is greatly simplified
if the distance $|\x_A-\x_B|$ is much larger than the sizes of wave packets 
of nuclei $A$ and $B$.  To take this into account, let us shift the variables 
of integration:

\begin{equation}
\label{coord}
\left\{
\begin{array}{rcl}
\x_1 & = & \x_A+\x_1' \\
\x_2 & = & \x_B+\x_2'  ~~.
\end{array}
\right.
\end{equation}

The wave packets of the nuclei $A$ and $B$ are essentially different from 
zero if $|\x_1'|$, $|\x_2'|  \ll  |\x_A-\x_B|$, hence

\begin{equation}
\label{n-vect}
|\x_1-\x_2| \simeq |\x_A-\x_B| + (\x_1'-\x_2')\n,
\end{equation}
where  $\n$ is the unit vector in the direction $\x_A-\x_B$.

By substituting eqs. (\ref{coord}) and (\ref{n-vect})  into eq. (\ref{Ampl'})
we obtain after integration over $d\x_1'$  and  $d\x_2'$:

\begin{multline}
\label{Ampl"}
\mathcal{A}_j \simeq \int
\delta \left( \sum_{n=e,A,B}E_n(q_n)-E_C-E_l-E_D \right)
e^{i(\q_e-\p_C) \x_A-i(\p_l+\p_D) \x_B-i\q_e \x_e+iE_e(\q_e)(t_e-t_A)}  
\times \\ \times 
e^{-\sum_{n=e,A,B}\frac{(\p_n-\q_n)^2}{2\sigma_n^2}}
\delta(\q_e+\q_A-k_j\n-\p_C) \delta(\q_B+k_j\n-\p_l-\p_D)
\times \\ \times
\frac{\exp{ \left\{ik_j|\x_A-\x_B| + i\omega_j (t_A-t_B) \right\}}}
{4\pi|\x_A-\x_B|} d\q_e d\q_A d\q_B
= \\ = 
\int \delta( \sum_{n'=A,B}E_{n'}(q_{n' j})+E_e(q_e) - E_C - E_l - E_D)
\frac{e^{ik_j|\x_A-\x_B| + i\omega(\q_{A j},\q_{B j}) \cdot (t_A-t_B)}}
{4\pi|\x_A-\x_B|}
\times \\ \times 
e^{i(\p_e-\p_C) \x_A - i(\p_l+\p_D) \x_B -i\p_e \x_e+iE_e (t_e -t_A)}
e^{-\sum_{n'=A,B}\frac{(\p_{n'}-\q_{n' j})^2}{2\sigma_{n'}^2} -
\frac{(\p_e-\q_e)^2}{2\sigma_e^2}} d\q_e.
\end{multline}

The $j$-dependent momenta $\q_{A j}$ and $\q_{B j}$ are defined as:
\begin{equation}
\label{q-q'-dep}
\begin{array}{rcl}
\q_{A j} & = & k_j\n + \p_C - q_{e j}\e ~,\\
\q_{B j} & = & \p_l + \p_D - k_j\n~.
\end{array}
\end{equation}

We integrate over $d\q_e$ using (\ref{dq_e}). The result is:
\begin{multline}
\label{Ampl_3}
\mathcal{A}_j \simeq |q_{e j}|^2
\left( \frac{dE_{\Sigma}(q_e)}{dq_e} \right)^{-1}
\frac{e^{ik_j|\x_A-\x_B| + i\omega_j (t_A-t_B)}}{4\pi|\x_A-\x_B|}
\times \\ \times 
e^{i(q_{e j} \e-\p_C) \x_A - i(\p_l+\p_D) \x_B -i\q_e \x_e + iE_e(\q_e)
(t_e-t_A)}
e^{-\sum_{n=e,A,B}\frac{(\p_n-\q_{n j})^2}{2\sigma_n^2}},
\end{multline}
where $\omega_j \equiv \omega(\q_{A j},\q_{B j})$ (see 
eq.~(\ref{omega-p-dep.})), 
$E_{\Sigma}(q_e) \equiv \sum_{n'=A,B}E_{n'}(q_{n' j})+E_e(q_e)-E_C-E_l-E_D$, 
and $\left( {dE_{\Sigma}(q_e)}/{dq_e} \right)^{-1}$ is the Jacobian, 
left after integration of the energy $\delta$-function in (\ref{Ampl"}). 
 (In eq.(\ref{Ampl"}) there are three $\delta$-functions, one of them 
expressing the energy conservation, while the other two refer to momentum 
conservation in $A$ and $B$ vertices.)
 
We have already stressed  that $k_j$ is not a momentum, but a parameter 
characterizing spherical neutrino wave. Now we see that in the case of very 
large distance $|\x_A-\x_B|$  the parameter $k_j \n$ does play the role of 
the neutrino momentum. We are faced with the situation when neutrino being 
virtual particle at short distance from the source becomes effectively real 
 at large distance, near detector.

\section{Phases of the amplitudes}

We are interested first of all in the phases of  amplitudes of the process
considered. From eq. (\ref{Ampl_3}) one can see that the phase of
$\mathcal{A}_j$ equals to

\begin{multline}
\label{phase}
\phi_j =  k_j|\x_A-\x_B| + \omega_j (t_A-t_B)
+ (q_{e j}\e-\p_C) \x_A 
- \\ - 
(\p_l+\p_D) \x_B - q_{e j}\e \x_e + E_e(p_{e j})(t_e-t_A),
\end{multline}
and  dependence of  $k_j$, $q_{e j}$ and $\omega_j$ on $m_j$ is given by 
the system of equations
 (\ref{omega-p-dep.}), (\ref{kappa}), (\ref{q-q'-dep}), and
 by the on-mass-shell conditions for nuclei and electron.

From eq. (\ref{phase}) it follows that
\begin{multline}
\label{phases}
\phi_{i j} \equiv \phi_i-\phi_j =  |\x_B-\x_A| (k_i - k_j) +
(\omega_i - \omega_j) (t_A-t_B) 
+ \\ +
(q_{e i}-q_{e j}) (\x_A-\x_e)\e 
+ (E_e(q_{e i})-E_e(q_{e j})) (t_e-t_A).
\end{multline}

In equation (\ref{phases}) the difference of the electron energies could be 
expressed through the difference of the corresponding momenta:
\begin{equation}
(E_e(q_{e i})-E_e(q_{e j})) = \frac{dE_e(q_e)}{dq_e} (q_{e i}-q_{e j}).
\end{equation}

Since neutrino masses are much smaller than energies and momenta of external
particles\footnote{
Let us point out that in the limit $M_{A,B} \rightarrow \infty$ the neutrino 
energy tends to:
$\omega_0 \rightarrow E_e + E_A - E_C = E_l - E_B + E_D = E_e + 
O({1}/{M_{A,B}})$, the value defined by the energy conservation
in the process $e + A + B \rightarrow l + C + D$, and since
 ${(\q\n)}/{E(\q)} \rightarrow 0$, the phase difference approaches 
its standard value
$$
\nonumber
\phi_{i j} \rightarrow  - |\x_B-\x_A|\frac{m^2_i-m^2_j}{2\omega} =
- \frac{m^2_i-m^2_j}{2E_e}|\x_B-\x_A| .
$$ } 
we may write:

\begin{equation}
\label{deltaK}
\delta k_{ij} \equiv
k_i - k_j \simeq (m^2_i-m^2_j) \cdot 
\left. \frac{dk_j}{d(m_j^2)} \right|_{m_j^2=0}
= -\frac{m^2_i-m^2_j}{2\omega_0 \cdot (1-\V_B \cdot \n)},
\end{equation}

\begin{equation}
\label{deltaOmega}
\delta \omega_{ij} \equiv
\omega_i - \omega_j \simeq  (m^2_i-m^2_j) \cdot 
\left. \frac{d\omega}{d(m_j^2)} \right|_{m_j^2=0}
= -\frac{(m^2_i-m^2_j)}{2\omega_0} \cdot 
\frac{\V_B \cdot \n}{1-\V_B \cdot \n}, 
\end{equation}

\begin{equation}
\label{deltaQe}
q_{e i} - q_{e j} \simeq  (m^2_i-m^2_j) \cdot
\left. \frac{d\q_{e j}}{d(m_j^2)} \right|_{m_j^2=0}
= -\frac{(m^2_i-m^2_j)}{2\omega_0 \cdot (1-\V_B\n)} \cdot 
\frac{(\V_B-\V_A)\n}{(\V_e-\V_A)\n},
\end{equation}
where $\V_n \equiv {\q_{n 0}}/{E_n(\q_{n 0})}$,
and the subscript $n$ denotes $e,~A,~B$.

The quantities  $\q_{n 0}$, $E_n(\q_{n 0})$ and $\omega_0$ are  defined 
by the external parameters from eqs.
(\ref{omega-p-dep.}), (\ref{kappa}), and (\ref{q-q'-dep}) at $m_j=0$.
In particular:
\begin{multline}
\label{omega0}
\omega_0 =
\frac{(\p_D-\p_l)^2+M_B^2-(E_D+E_l)^2}{2[(\p_D + \p_l)\n-E_D-E_l]} 
\simeq
\frac{(\p_C-\p_e)^2+M_A^2-(E_C-E_e(\p_e))^2}
{2[(\p_e-\p_C)\n+E_C-E_e(\p_e)]}.
\end{multline}

In eq. (\ref{omega0}) the sign "=" means exact but somewhat useless 
equality, because $\p_D$ and $E_D$ are not measured, while $\p_l$ is 
measured with low accuracy. As for the sign "$\simeq$", it will be used in 
what follows because $\p_e$ and $E_e$ are known and essentially define the 
value of $\omega_0$:
\begin{equation}
\label{omega0=E_e}
\omega_0 = E_e + O(\frac{1}{M_A})~.
\end{equation}

From eqs. (\ref{phases}-\ref{deltaQe}), 
it follows that
\begin{multline}
\label{phi_ij}
\phi_{i j} = \frac{m_i^2-m_j^2}{2\omega_0 (1-\V_B \cdot \n)}
\times  \\ \times
 \left( -|\x_B-\x_A| + (t_B-t_A) \V_B \cdot \n +
 [(\x_e-\x_A) \e - |\V_e|(t_e-t_A)] \cdot 
\frac{(\V_B-\V_A)\n}{(\V_e-\V_A)\n} \right)~.
\end{multline}

It is convenient to choose the parameters $t_e$, $t_A$, $\x_e$, 
and $\x_A$ in such a way that $\x_e = \x_A$ when $t_e = t_A$. This convention 
corresponds to the classical picture of $e A$-collision and allows to 
simplify eq. (\ref{phi_ij}):
\begin{equation}
\label{SimplePhi_ij}
\phi_{i j} = \frac{m_i^2-m_j^2}{2\omega_0 (1-\V_B \cdot \n)}
 \left(- |\x_B-\x_A| + (t_B-t_A) \V_B \cdot \n \right).
\end{equation}

From eqs.(\ref{SimplePhi_ij}), (\ref{deltaK}) and (\ref{deltaOmega})
it follows:

\begin{equation}
\label{phase-end}
\phi_{ij} = \delta k_{ij} |{\bf x}_B - {\bf x}_A | -
\delta \omega_{ij} (t_B - t_A ) =
\end{equation}

$$
=
-\frac{m_i^2-m_j^2}{2\omega_0 } |{\bf x}_B - {\bf x}_A | + \delta \omega_{ij}
\left [|{\bf x}_B - {\bf x}_A | -( t_B - t_A) \right ] .
$$
The first term in the phase is the standard phase of oscillation theory,
while the second one is an additional term which depends upon the size 
of the electron wave packet.

\section{Neutrino oscillations and their suppression}

By using eq. (\ref{Ampl'}) we obtain the following expression 
for $P_{i j}$ describing neutrino oscillations in the right-hand side 
of eq. (\ref{P_{exp}}):
\begin{multline}
\label{SupprProb}
P_{i j} = e^{-i\phi_{i j}}
\frac{|q_{e i}|^2|q_{e j}|^2}{16\pi^2|\x_B-\x_A|^2}
\left( \frac{dE_{\Sigma}(q_e)}{dq_e} \right)^{-2}
\times \\ \times
\exp \left(-\sum_{n=e,A,B}\frac{(\p_n-\q_{n i})^2}{2\sigma_n^2}
-\sum_{n=e,A,B}\frac{(\p_n-\q_{n j})^2}{2\sigma_n^2} \right)~,
\end{multline}
where $\phi_{i j}$ is given by eq. (\ref{SimplePhi_ij}).

This formula allows to compare the oscillating terms ($i \neq j$) with 
non-oscillating
($i = j$) ones, and thus to analyze the strength of oscillations as a 
function of the momentum spread of the electron wave packet.

For easier comparison with ref. \cite{KMOS}  we assume 
in what follows that $\sigma_e$ is much smaller than $\sigma_A$ and 
$\sigma_B$. 
Let us define
\begin{equation}
\label{fi}
f_i \equiv \exp(-\frac{(\p_e-\q_{e j})^2}{2\sigma_e^2})
\end{equation}
and assume that $|\p_e-\q_{e 1}| < |\p_e-\q_{e 2}|, |\p_e-\q_{e 3}|$, 
then $f_1 \gg f_2, f_3$ in the limit of vanishing $\sigma_e$. 
The leading diagonal term
\begin{equation}
\label{P33}
P_{1 1} \sim f_1^2 \gg P_{2 2}, P_{3 3}.
\end{equation}
Comparing $P_{1 1}$ with the non-diagonal terms we conclude:
\begin{equation}
\label{P11}
P_{1 1} \sim f_1^2 \gg P_{1 2} \sim f_1 \cdot f_2,~~~
\mathrm{and}~~~
P_{1 1} \gg P_{1 3}, P_{2 3}.
\end{equation}
Considering the ratio
\begin{equation}
\label{fifj}
\frac{f_j}{f_i} = \exp \left( 
- \frac{(\p_e - \frac{\q_{e j}+\q_{e i}}{2})}{\sigma_e^2}
\cdot (\q_{e j} - \q_{e i}) \right)
\end{equation}
and using eq. (\ref{deltaQe}) one finds the crucial parameter of 
suppression to be 
\begin{equation}
\label{Lij}
\frac{q_{e j} - q_{e i}}{\sigma_e} = 
\frac{(\V_B-\V_A) \cdot \n}{\sigma_e \cdot L_{i j}},
\end{equation}
where
\begin{equation}
\label{Lij-1}
L_{i j}^{-1} \equiv \frac{(m^2_i-m^2_j)}{2\omega_0}
\end{equation}
is $i j$ oscillation length.

The Gaussian factor in eq.(\ref{SupprProb}) makes it obvious that 
for $|\q_{n i}-\q_{n j}| \gg \sigma_n$, the oscillating terms become 
exponentially suppressed in comparison to the non-oscillating ones.

In conclusion of this section let us make the following remark. Though the 
suppression for vanishing $\sigma_e$ is obvious, it is clear, that in 
"a realistic thought experiment" $\sigma_e^{-1} \ll L_{osc}$, and hence 
the suppression is  very weak.

\section{Concluding remarks}  

1) We see that the alternative "equal energies versus equal momenta" is 
naturally resolved if one consistently uses the standard rules
of quantum mechanics and in particular quantum field theory.
In the example, which we consider here,  
using the propagator of virtual neutrinos and mixed description
of initial (wave packets) and final (plane waves)  particles 
 all kinematical variables are uniquely defined.
In particular, when we go beyond plane wave approximation
for initial particles there is no equal momenta nor equal energies. However, 
still $|\omega_j -\omega_i| \ll |k_j-k_i|$ at least for
non-relativistic nuclei. 
Similar conclusions were obtained in refs. \cite{erice}, \cite{ad-rev} 
for oscillating neutrinos produced in pion decay 
(see also \cite{MN}). \\
2) For the plane wave of the initial electron and finite mass nuclei, 
neutrino oscillations disappear unlike the case of infinitely heavy nuclei. 
For a finite but small momentum spread of the electron wave packet, the 
neutrino oscillations are suppressed.  \\
3) For realistic parameters of the electron wave packet the above suppression 
is small and therefore can be disregarded.  \\
4) The Green function used to describe neutrino leads us to the situation 
when the neutrino being a virtual particle at short distances from the 
source, becomes effectively a real particle at large distances, near 
detector. This is the standard case in the scattering theory. \\
5) With localized "meeting points" $eA$ and $lB$ the time dependence of 
the oscillation probability is not essential. (The time moments $t_A$ and 
$t_B$  enter the expression for $\phi_{i j}$ with small coefficients 
proportional to velocities of nonrelativistic nuclei.)

\section{Acknowledgments}
We are grateful to M. Vysotsky for valuable comments. The work was partly 
supported by RFBR grants No. 2328.2003.2, No. 04-02 - 16538, by INTAS
grant 00561 and by the A. von Humboldt award to L.O.

\end{document}